%% file: main.tex
\pdfoutput=1

\documentclass[10pt,twocolumn,letterpaper]{article}

\usepackage[pagenumbers]{cvpr} 

\usepackage{graphicx}
\usepackage{amsmath}
\usepackage{amssymb}
\usepackage{booktabs}
\usepackage{epsfig}
\usepackage{enumitem}
\usepackage[ruled,vlined]{algorithm2e}
\usepackage{multirow}
\usepackage{color,soul}
\usepackage{float}
\newcommand{\name}{ExPLoit}

\usepackage[pagebackref,breaklinks,colorlinks]{hyperref}

\usepackage[capitalize]{cleveref}
\crefname{section}{Sec.}{Secs.}
\Crefname{section}{Section}{Sections}
\Crefname{table}{Table}{Tables}
\crefname{table}{Tab.}{Tabs.}

\def\cvprPaperID{6170} 
\def\confName{CVPR}
\def\confYear{2022}

\begin{document}

\title{\name: \underline{Ex}tracting \underline{P}rivate \underline{L}abels in Spl\underline{it} Learning}
\author{Sanjay Kariyappa\\
Georgia Institute of Technology\\
Atlanta GA, USA\\
{\tt\small sanjaykariyappa@gatech.edu}
\and
Moinuddin K Qureshi\\
Georgia Institute of Technology\\
Atlanta GA, USA\\
{\tt\small moin@gatech.edu}
}
\maketitle

\newcommand{\uptox}{99.96}

\newcommand{\ignore}[1]{}

\begin{abstract}
\emph{Split learning} is a popular technique used for vertical federated learning (VFL), where the goal is to jointly train a model on the private input and label data held by two parties. This technique uses a split-model, trained end-to-end, by exchanging the intermediate representations (IR) of the inputs and gradients of the IR between the two parties. We propose \emph{ExPLoit} -- a label-leakage attack that allows an adversarial input-owner to extract the private labels of the label-owner during split-learning. ExPLoit frames the attack as a supervised learning problem by using a novel loss function that combines gradient-matching and several regularization terms developed using key properties of the dataset and models. Our evaluations show that ExPLoit can uncover the private labels with near-perfect accuracy of up to $\uptox \%$. Our findings underscore the need for better training techniques for VFL.
\end{abstract}


\input{1.introduction}
\input{2.related}
\input{3.preliminaries}
\input{4.our_proposal}
\input{5.experiments}

\input{6.gradient_noise}
\input{7.conclusion}
\input{acknowledgements}
{\small
\bibliographystyle{ieee_fullname}
\bibliography{references}
}
\clearpage
\input{8.appendix}

\end{document}

%% file: 1.introduction.tex
\section{Introduction}
\label{sec:intro}

\begin{figure*}[htb]
	\centering
    \centerline{\epsfig{file=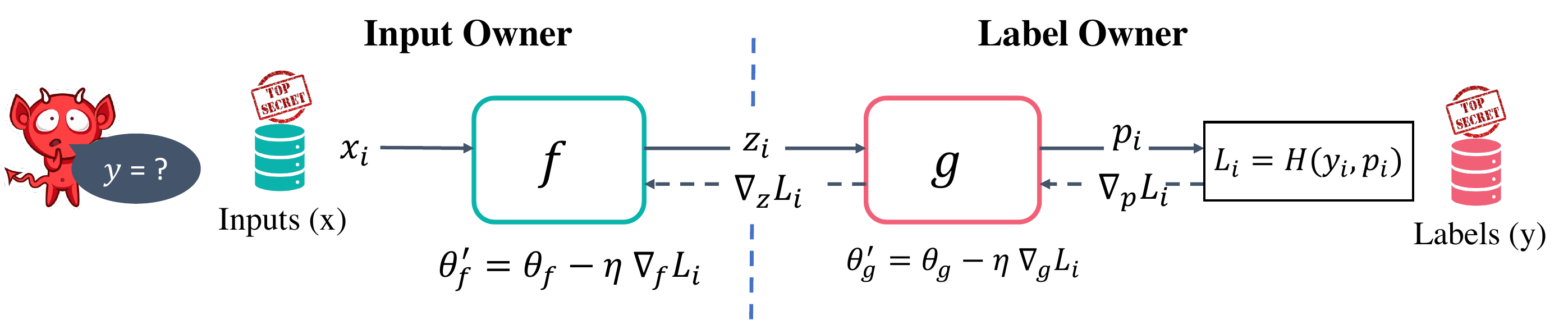, width=0.8\textwidth}}
	\caption{Split learning can be used for vertical federated learning by training a composition model $g\circ f$, split between the input and model owner. 
	In this work, we demonstrate that an adversarial input owner can learn the private labels using gradient information obtained during split learning, compromising the label owner's privacy.}
    \label{fig:split_learning}
\end{figure*}

The plethora of apps and services that we use for our everyday needs, such as online shopping, social media, communication, healthcare, finance, etc., have created distributed silos of data. While aggregating this distributed data would improve the performance of machine learning models, doing so is not always feasible due to privacy constraints. For instance, in healthcare, laws like HIPAA require hospitals to keep medical records private. For finance and internet companies, user agreements and privacy laws might prevent them from sharing data. These challenges have led to the development of several techniques for \emph{federated learning}, which allow models to be trained without the data owner having to share their data explicitly. \emph{Split learning} ~\cite{sl1,sl2} is one such technique that allows federated learning to be performed when the inputs and the corresponding labels are held by two different parties. Split learning uses two models $f:\mathcal{X}\rightarrow\mathcal{Z}$ and $g:\mathcal{Z}\rightarrow\mathcal{Y}$ that are split between the input and label owners as shown in Fig.~\ref{fig:split_learning}. The composition network $g\circ f$ is trained end-to-end, with the input owner transmitting the embedding $z_i$ (intermediate representation) to the label owner in the forward pass and the label owner returning the gradient $\nabla_zL_i$ to the input owner during the backward pass. This allows the split model to be trained on the distributed data while keeping the sensitive data with their respective owners.

Unfortunately, split learning does not have formal privacy guarantees, and it is not clear if it allows the input and label owners to hide their private data from each other. We set out to answer this question by considering an adversarial input owner who wants to break label privacy by extracting the private labels in two-party split learning. To this end, we propose \emph{ExPLoit}-- a label-leakage attack that frames the problem of learning the private labels as a supervised learning task, by leveraging the gradient information ($\nabla_zL_i$) obtained during split learning. ExPLoit ``replays" split learning by replacing the label owner's model $g$ and labels $\{y_i\}$ with a randomly initialized surrogate model $g'$ (with parameters $\theta_{g'}$) and surrogate labels $\{y'_i\}$ respectively. We aim to learn the private labels by training these surrogate parameters using the following key objectives:
\begin{enumerate}
    \item \emph{Gradient Matching Objective:} The gradient computed using the surrogate model and labels during \emph{replay split learning} should match the gradients received from the label owner during the original split learning process. 
    \item \emph{Label Prior Objective:} The distribution of surrogate labels must match the expected label prior distribution. For instance, if the classification problem in consideration has a uniform label prior, the surrogate labels must also have a uniform distribution.
    \item \emph{Label Entropy Objective:} Since we consider datasets with hard labels, each individual surrogate label $y'_i$ must have low entropy.
    \item \emph{Accuracy Objective:} The predictions made by the surrogate model must be close to the surrogate labels, achieving high accuracy. 
\end{enumerate}

Combining the above objectives yields a loss function that can be used to train the surrogate parameters and labels. By minimizing this loss function over all the embedding, gradient pairs $\{z_i, \nabla_zL_i\}$ received during split learning, the surrogate labels $\{y'_i\}$ can be trained to match the private labels of the label owner $\{y_i\}$, allowing an adversarial input owner to carry out a label leakage attack with high accuracy. Our paper makes the following key contributions:

\textbf{Contribution 1:} We propose \emph{ExPLoit} -- a label-leakage attack for two-party split learning. 
Our proposal replaces the label owner's private labels and model with surrogate parameters and frames the attack as a supervised learning problem. We combine gradient-matching and regularization terms obtained by leveraging several key properties of the model and dataset to develop a novel loss function, which can be used to train the surrogate parameters and uncover the private labels of the label owner.

\textbf{Contribution 2:}  We carry out extensive evaluations on the Criteo conversion prediction~\cite{criteo} task, and several image classification datasets, including MNIST, FashionMNIST, CIFAR-10, and CIFAR-100 to show that ExPLoit can leak private labels with near-perfect accuracy (up to $\uptox \%$) for most datasets. Our attack is effective across multiple model architectures and also outperforms several recently proposed label-leakage attacks for split learning.

\textbf{Contribution 3:} We evaluate perturbing the gradient with noise as a defense against our attack. Our results show that gradient noise allows the label owner to trade off model accuracy (lower utility) for improved privacy against ExPLoit (better label privacy). While this technique works well for simpler datasets like MNIST, it leads to significant degradation in accuracy for datasets like CIFAR-10 and CIFAR-100, making it unsuitable for more complex datasets.

Our findings in this work demonstrate that split learning does not protect the privacy of labels, emphasizing the need for better techniques for VFL.

%% file: 2.related.tex
\section{Related Work}
We discuss prior work on label leakage attacks and describe their limitations. For an overview of techniques besides split learning that can be used to learn on vertically partitioned data, we refer the reader to Appendix~\ref{app:other_related}.

\begin{figure}[htb]
	\centering
    \centerline{\epsfig{file=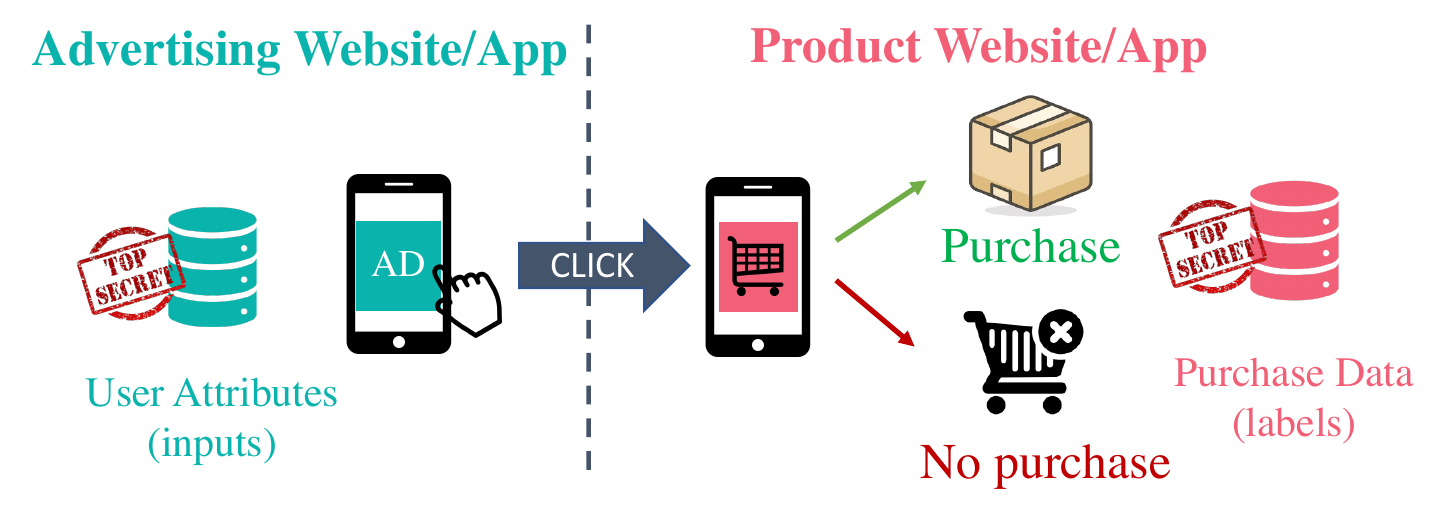, width=\columnwidth}}
	\caption{Conversion prediction estimates the likelihood of a purchase when a user clicks on an ad. The training data is vertically partitioned, with the user attributes (inputs) held by the ad company and the purchase data (outputs) held by the product company (Figure adapted from ~\cite{conversion_prediction_fig}).}
    \label{fig:conversion_prediction}
\end{figure}

\subsection{Norm-Based Attack for Conversion Prediction}  

Recently, Li et al. proposed \emph{Norm-Based Attack}~\cite{label_leakage} -- a label leakage attack on two-party split learning, specifically for the conversion prediction problem. We first provide background on the conversion prediction problem and then describe the attack.

\textbf{Conversion Prediction:} Given the attributes of a user and an ad, conversion prediction estimates the likelihood of a user purchasing the product. Conversion prediction is an essential component of ad-ranking algorithms, as ads with a high likelihood of conversion are more relevant to the user and need to be ranked higher. The data required to train the model are split between the advertising and product websites, as depicted in Fig.~\ref{fig:conversion_prediction}. The user attributes, which serve as the inputs, are stored with the advertising company, while the purchase data, which serve as the labels, are held with the product company. The companies are interested in training a model to predict the conversion likelihood while keeping their datasets private.

\textbf{Norm-based Attack:} Norm-based attack~\cite{label_leakage} leverages the observation that only a small fraction of ad clicks result in a purchase. Consequently, there is a high class imbalance in the training dataset of the conversion prediction task. This imbalance results in the magnitude of the gradients being higher when the infrequent class is encountered. Thus, by considering the norm of the loss gradient ($\|\nabla_zL_i\|_2$), an adversarial input owner can infer the private labels. Note that a key limitation of this attack is that it only works on binary classification problems with high class imbalances. In contrast, our proposed ExPLoit attack does not require a class imbalance and works for multi-class classification problems.

\subsection{UnSplit: Gradient Matching Attack}
Similar to our attack, a recent concurrent work \emph{UnSplit}\cite{unsplit} also proposes to learn the private labels in split learning using a gradient-matching objective~\cite{dlg} by minimizing the mean squared error (MSE) between the surrogate and true gradients using the following objective: $\min_{\theta_g', \{y'_i\}} MSE(\nabla_{z}L'_i, \nabla_{z}L_i)$. Here, $\theta_g'$ are the parameters of the surrogate model and $ \{y'_i\}$ are the surrogate labels. Results from this work show that UnSplit only works well when the label-owner's model $g$ is one-layer deep. In contrast, ExPLoit provides high accuracy of up to $\uptox\%$, even when the label owner uses multi-layer networks (up to 8 layers deep in our experiments). This is because ExPLoit uses additional regularization terms in the loss that help avoid poor local minima during training.

\subsection{Model Completion Attack}

Model Completion Attack~\cite{lia_vfl} uses unlabeled embeddings  $\mathcal{D} = \{z_i\}$ and a small number of labeled embeddings $\mathcal{D}^l = \{z^l_i, y^l_i\}$ to train a surrogate model $g':\mathcal{Z}\rightarrow\mathcal{Y}$ using semi-supervised learning. Since $g'$ functionally approximates the label-owner's model $g$, it can be used to predict the labels for the input embeddings $y'_i = g'(z_i)$, allowing the input owner to guess the private labels. This proposal suffers from two key drawbacks. First, it requires the adversary to have access to labeled examples, which may not always be available. For instance, in case of conversion prediction, labels for the input data cannot be gathered even using human annotators, as it is not readily apparent from the data. Second, the efficacy of this attack is limited by the accuracy of the model that can be trained by the adversary. Thus, the attack accuracy is highly dependent on the number of labeled examples available to the attacker and the difficulty of the prediction problem at hand. For example, this attack provides a label leakage accuracy of $91.46\%$ for MNIST, where a high-accuracy $g'$ can be trained with just a few labeled examples. However, the accuracy is much lower ($15.3\%$) for CIFAR-100, where the surrogate model is harder to train. In contrast, ExPLoit does not require any labeled examples and can achieve a high accuracy of $94.38\%$ even for complex datasets like CIFAR-100 (after 10 training epochs of the split-model).

%% file: 3.preliminaries.tex
\section{Preliminaries}
In this section, we provide background on the two-party split learning framework and formally state the objectives of the label leakage attack and defense.

\subsection{Two-Party Split Learning}


Two-party split learning is used for VFL, where the data is vertically partitioned between the input and label owner. The input owner owns the inputs $\mathcal{D}_{inp} = \{x_i\}$ and the label owner owns the labels $\mathcal{D}_{label} = \{y_i\}$, corresponding to each input. The goal of split learning is to train a composition model $g\circ f$ that is distributed between the two parties. Training with supervised learning requires mapping the entries in the input set to the corresponding entries in the label set. If this mapping is not known, private set intersection algorithms~\cite{psi} can be used to link the corresponding entries in the two datasets. A single training iteration involves a forward and a backward pass (as shown in Fig.~\ref{fig:split_learning}), which proceeds as follows:
\begin{itemize}
\item\emph{Forward pass:} The input owner samples a batch of inputs $\{x\}_{batch} \sim \mathcal{D}_{inp}$ and performs forward propagation through $f:\mathcal{X}\rightarrow\mathcal{Z}$ and produces the corresponding embeddings $\{z\}_{batch}$. These embeddings, along with the corresponding \emph{inputIDs} are sent to the label owner. The label owner feeds the embeddings to $g:\mathcal{Z}\rightarrow\mathcal{Y}$ to produce the predictions $\{p\}_{batch}$, which along with the labels $\{y\}_{batch}$ are used to compute the model's loss $L= \mathbb{E}[H(y, p)]$.

\item\emph{Backward pass:} The label owner initiates backpropagation and returns the loss gradient $\{\nabla_zL\}_{batch}$ to the input owner. Both the label and input owner compute the gradient of the loss with respect to the model parameters and update model parameters using gradient descent as shown below:
\begin{align}
\theta_g^{t+1} = \theta_g^{t} - \eta\nabla_{\theta_g}L;\ \ \theta_f^{t+1} = \theta_f^{t} -  \eta\nabla_{\theta_f}L.
\end{align}
\end{itemize}

\textbf{Privacy Objectives:} There are two key privacy objectives that split learning aims to achieve:
\begin{enumerate}
\item \emph{Input privacy:} The label owner should not be able to infer the input owner's private inputs $\{x_i\}$.

\item \emph{Label privacy:} The input owners should not be able to infer the label owner's private labels $\{y_i\}$.
\end{enumerate}

\subsection{Label Leakage Attack Objective}
In this work, we propose a label leakage attack, where an adversarial input owner tries to learn the label owner's private labels $\mathcal{D}_{label}=\{y_i\}$. We consider an honest-but-curious adversary, where the input owner tries to infer the private labels while honestly following the split learning protocol. During the training process of split learning, for each input $x_i$, the adversarial input owner transmits the embeddings $z_i$ and receives the gradient $\nabla_zL_i$ from the label owner. The adversary uses an algorithm $\mathcal{A}$ to estimate the private labels $y'_i$ for each input using these $\{z_i, \nabla_zL_i\}$ pairs obtained during split learning as shown below:
\begin{align} \label{eq:pseudo_labels}
\mathcal{A}(\mathcal{D}_{grad}) \rightarrow \mathcal{D}_{y'},\ \ \text{where } \mathcal{D}_{grad} = \{z_i, \nabla_zL_i\},\  \mathcal{D}_{y'} = \{y'_i\}.
\end{align}
The attack objective is to maximize the accuracy of estimated labels as follows:
\begin{align} \label{eq:adv_obj}
\max\mathop{\mathbb{E}}_{y_i\sim \mathcal{D}_{label}}[Acc(y_i, y'_i)].
\end{align}
Since the private labels $\{y_i\}$ are unavailable to the input owner, evaluating Eqn.~\ref{eq:adv_obj} is not possible. Instead, our proposed attack uses a surrogate objective that can be optimized to uncover the private labels with high accuracy.

\subsection{Label Leakage Defense Objective}
Defending against label leakage attack requires balancing two objectives:

\textbf{Utility Objective:} Train the composition model $g\circ f$ to have high classification accuracy on an unseen validation set (Eqn.~\ref{eq:defense_utility_obj}). 
\begin{align} \label{eq:defense_utility_obj}
\max\mathop{\mathbb{E}}_{x_i,y_i\sim \mathcal{D}_{val}}[Acc(y_i, g(f(x_i))]
\end{align}

\textbf{Privacy Objective:} Minimise the accuracy of the estimated labels $\{y'_i\}$ that can be recovered from the gradient information (Eqn.~\ref{eq:defense_privacy_obj}).
\begin{align} \label{eq:defense_privacy_obj}
\min\mathop{\mathbb{E}}_{y_i\sim \mathcal{D}_{label}}[Acc(y_i, y'_i)]
\end{align}

%% file: 4.our_proposal.tex
\section{Our Proposal: ExPLoit}
We propose \emph{ExPLoit}-- a label leakage attack that can be used by a malicious input-owner to learn the private labels in split learning. Our key insight is that the label leakage attack can be framed as a supervised learning problem by replacing the unknown parameters of the label owner with learnable surrogate variables. This allows the adversarial input owner to ``replay" the split learning process with surrogate variables. We develop a novel loss function using gradient-matching and several regularization terms developed using properties of the model and training data. By minimizing this loss function, we can recover the label owner's private labels with high accuracy. The rest of this section describes our proposed attack in greater detail.

\begin{figure*}[htb]
	\centering
    \centerline{\epsfig{file=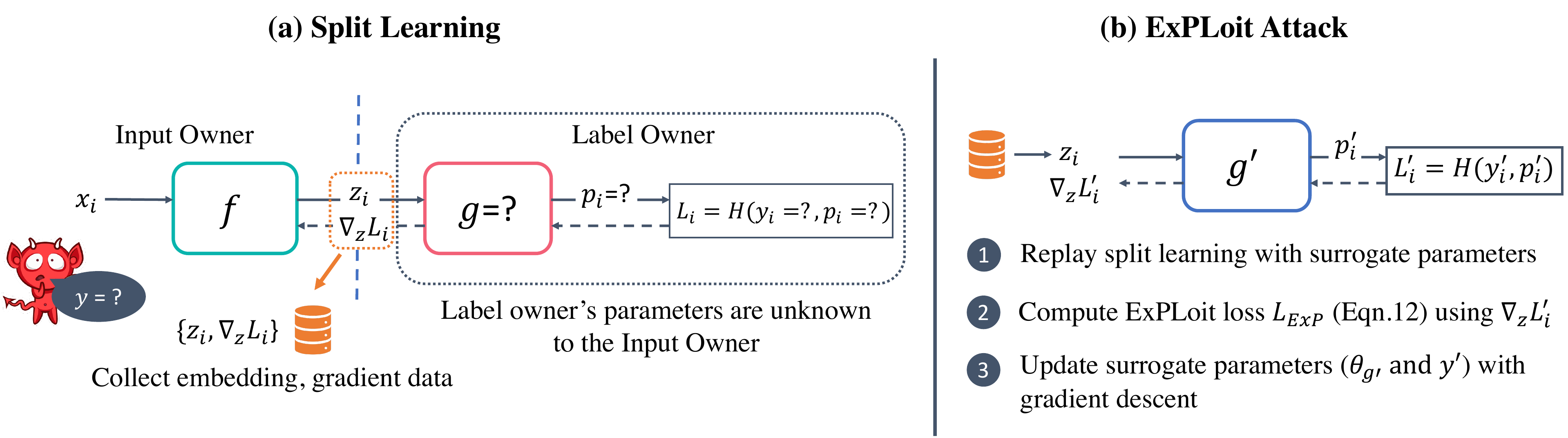, width=\textwidth}}
	\caption{ (a) \emph{Split Learning:} The adversarial input owner collects the embedding and gradient data $\{z_i, \nabla_zL_i\}$ when performing split learning with the label owner. (b) \emph{ExPLoit Attack:} The embedding and gradient data is used to train surrogate model and label parameters ($g'$ and $\{y'_i\}$) and uncover the private labels.}
    \label{fig:learning_problem}
\end{figure*}

\subsection{Surrogate Variable Substitution} 

From the input owner's point of view, the split learning process has two key unknowns: the label owner's model $g$ and the private labels $\{y_i\}$ (see Fig.~\ref{fig:learning_problem}a). Our goal is to uncover these unknown values by treating them as learnable parameters. To do so, we start by substituting these unknowns with randomly initialized surrogate parameters, as shown in Fig.~\ref{fig:learning_problem}b. We replace $g$ with a surrogate model $g'$ (with parameters $\theta_{g'}$), and $\{y_i\}$ with a set of surrogate labels $\{y'_i\}$. We want $y'_i$ to be a point on an $n-1$ dimensional probability simplex for an n-class classification problem. To enforce this property, we set $y'_i = Softmax(\hat{y}_i)$, where $\hat{y}_i\in\mathcal{R}^n$. With the surrogate parameters in place, the goal of our attack is to learn the surrogate labels $\{y'_i\}$ (or equivalently to learn $\{\hat{y}_i\}$).

\subsection{Replay Split Learning} 

To train the surrogate parameters, we first ``replay" the split learning process using the surrogate variables (Fig.~\ref{fig:learning_problem}b). First, in the forward pass, the embedding $z_i$ (collected during split learning) is fed into $g'$ to get the prediction $p'_i$, which along with the surrogate labels $y'_i$ can be used to compute the loss $L'_i=H(y'_i, p'_i)$. Next, we perform backpropagation through $g'$ and compute the gradient of the loss with respect to the embedding $\nabla_{z}L'_i$. We use this gradient data as part of our loss function to learn the private labels, as described below.


\subsection{ExPLoit Loss}
To train the surrogate parameters $\theta_{g'}$ and $\hat{y}$, we formulate a loss function using four key objectives:

1. \emph{Gradient Objective}: The loss gradient $\nabla_zL'_i$, obtained during \emph{replay split learning}, must match the original gradients $\nabla_{z}L_i$, obtained during the original split learning process. This can be achieved by minimizing the $l^2$ distance between $\nabla_{z}L'_i$ and $\nabla_{z}L_i$ as shown below:
\begin{align} \label{eq:grad_objective}
\min_{\theta_g', \{\hat{y}_i\}} \mathbb{E} \left\|\nabla_{z}L'_i - \nabla_{z}L_i \right\|_2.
\end{align}

2. \emph{Label Prior Objective}: The distribution of surrogate labels must match the label prior $P_y$ of the dataset\footnote{We assume that the input owner knows the label prior distribution (see Appendix~\ref{app:limitations}).}. The probability distribution of the surrogate labels can be computed by taking the expectation of the surrogate labels\footnote{Each surrogate label $y'_i$ represents a probability distribution over the output classes.} $P_{y'} = \mathbb{E}(y'_i)$. We perform the following optimization to match the distributions of the original and surrogate labels:
\begin{align} \label{eq:label_prior_objective}
\min_{\theta_g', \{y'_i\}}\mathcal{D}_{KL}(P_{y}\| P_{y'}).
\end{align}

3. \emph{Label Entropy Objective}: Each individual surrogate label $y'_i$ must have low entropy as the datasets we consider have zero entropy one-hot labels. 

4. \emph{Accuracy Objective}: The surrogate model must have high prediction accuracy with respect to the surrogate labels. In other words, the predictions of the surrogate model $p'_i$ must be close to the surrogate labels $y'_i$.

To achieve the label entropy and accuracy objectives, we can minimize the normalized cross-entropy loss between $p'_i$ and $y'_i$ as follows:
\begin{align} \label{eq:accuracy_objective}
\min_{\theta_g', \{y'_i\}_i}\frac{\mathbb{E}[H(y'_i, p'_i)]}{H(P_y)}. 
\end{align}
Note that the cross-entropy term $H(y'_i, p'_i)$ in Eqn.~\ref{eq:accuracy_objective} can be expressed as a sum of the label entropy and KL divergence between the surrogate label and prediction: $H(y'_i, p'_i) = H(y'_i) + \mathcal{D}_{KL}(y'_i\| p'_i)$. Thus, by minimizing cross-entropy, we can minimize the entropy of surrogate labels (label entropy objective) and match the model's predictions with the surrogate labels (accuracy objective). We normalize cross-entropy with the entropy of the label prior $H(P_y)$ to ensure that the metric is insensitive to the number of label classes and the label priors~\cite{norm_ce}.

We combine all the learning objectives described above to derive the final loss function as shown below:
\begin{align}\label{eq:loss}
L_{ExP} &= \mathbb{E}\Big[ \left\|\nabla_{z}L_i - \nabla_{z}L'_i\right\|_2\Big] +\\ \nonumber &\lambda_{ce}\cdot\mathbb{E}\Big[H(y'_i, p'_i)/H(P_y)\Big] + \lambda_p\cdot\mathcal{D}_{KL}(P_{y}\| P_{y'}).
\end{align}
Here, $\lambda_{ce}$ and $\lambda_p$ dictate the relative importance of the cross-entropy and label prior terms compared to the gradient loss term (first term in Eqn.~\ref{eq:loss}). By optimizing the surrogate model and label parameters using this loss function, we can recover the private labels of the label owner with high accuracy. We consider the gradient loss term to be the primary optimization objective of our loss function. The cross-entropy and label prior terms act as regularizers and help us achieve a better label leakage accuracy (see Appendix~\ref{app:ablation} for an ablation study).

\subsection{Putting It All Together}
The individual components described thus far can be combined to carry out our label leakage attack. Our attack starts with the input owner performing split learning process with the label owner, as shown in Fig.~\ref{fig:learning_problem}a. During this process, the input owner collects the embedding $z_i$ and the corresponding loss gradient $\nabla_{z}L_i$ for each input. Using this data, the input owner can use the \emph{ExPLoit} attack to leak the private labels. Our attack is described in Algorithm~\ref{alg:attack}. In the outer loop, we pick values for $\lambda_p, \lambda_{ce}$ and the learning rates $\eta_{g'}, \eta_{\hat{y}}$ using a Bayesian hyperparameter optimization algorithm. The surrogate parameters $\{\hat{y}_i\}_i$ and $\theta_{g'}$ are randomly initialized, and each inner loop of the attack proceeds as follows:
\begin{enumerate}
\item Replay split learning with surrogate parameters with the following steps:
\begin{enumerate}[label=\alph*.]
    \item Sample a batch of embeddings, gradients and surrogate labels: \linebreak $\{z, \nabla_zL, \hat{y}\}_{batch}$.
    \item Perform forward pass and compute loss $\{L'\}$ using predictions $\{p'\}$ and surrogate labels $\{y'\}$.
    \item Perform backpropagation to compute the loss gradients $\{\nabla_{z}L'\}$.
\end{enumerate}
\item Compute the ExPLoit loss: $L_{ExP}$ (Eqn.~\ref{eq:loss}).
\item Update surrogate parameters $\theta_{g'}$ and $\{\hat{y}_i\}$ to minimize $L_{ExP}$.
\end{enumerate}
We repeat the above steps until the surrogate parameters converge.
\begin{algorithm}[h]
\SetAlgoLined 
\KwIn{$\{z_i\}, \{\nabla_{z}L_i\}, P_{y}, N_{iter}$}
\KwOut{$\{y^{*}_i\}$}
 
 $\mathcal{D}_{train} = \{z_i, \nabla_{z}L_i, y'_i\}$\\
 
\For{$i\gets0\ to\ N_{iter}$}{ 
$\lambda_p, \lambda_{ce}, \eta_{g'}, \eta_{\hat{y}} \gets BayesOpt()$\\
Initialize $\{\hat{y}_i\}, g'(\cdot;\theta_{g'})$\\
\Repeat{Convergence}{

\For{$\{z, \nabla_{z}L, \hat{y}\}_\text{batch}$ in $\mathcal{D}_\text{train}$} {
\texttt{\\}
    $\{y'_i\} = \{Softmax(\hat{y}_i)\}$\\
    $P_{y'} = \mathbb{E}(y'_i)$\\
    \texttt{\\}
    // 1. Replay Split Learning\\
    \For{$\{z, \nabla_{z}L, \hat{y}\}_\text{i}$ in $\{z, \nabla_{z}L, \hat{y}\}_\text{batch}$} {
    
    $p'_i = g'(z_i; \theta_{g'})$\\
    
    $L'_i = \mathcal{D}_{KL}(y'_i\| p'_i)$\\
    Compute $\nabla_{z}L'_i$\\
    }

\texttt{\\}
// 2. Compute ExPLoit loss (Eqn. 12)\\
$L_{ExP} =  \mathbb{E}[\left\|\nabla_{z}L'_i - \nabla_{z}L_i\right\|_2] + \lambda_{ce} \cdot \mathbb{E}[H(y'_i, p'_i)/H(P_y)] + \lambda_p\cdot \mathcal{D}_{KL}(P_{y}\| P_{y'}) $\

\texttt{\\}
// 3. Update surrogate model, label parameters\\
$\theta_{g'} \gets \theta_{g'} - \eta_{g'}\cdot \nabla_{\theta_{g'}}L_{ExP}$\\
$\hat{y} \gets \hat{y} - \eta_{\hat{y}} \cdot \nabla_{\hat{y}}L_{ExP}$\\

}
}
$NewBest = UpdateBayesOpt(\mathbb{E}[\left\|\nabla_{z}L'_i - \nabla_{z}L_i\right\|_2])$ \\
\If{$NewBest$}{ 
    $\{y^{*}_i\} \gets \{Softmax(\hat{y}_i)\}$
}
}
\caption{ExPLoit Attack}
\label{alg:attack}
\end{algorithm}

\textbf{Hyperparameter Optimization}: We learn a set of surrogate labels $\{y'_i\}$ in each outer loop for different selections of the hyperparameters. Unfortunately, evaluating the accuracy of the surrogate labels produced in each iteration is not possible since the input owner is unaware of any of the true labels. Consequently, we cannot use accuracy to guide the hyperparameter search. Instead, we evaluate the gradient loss term $\mathbb{E}[\left\|\nabla_{z}L'_i - \nabla_{z}L_i\right\|_2]$ after completing each outer iteration and use this as our objective function to be minimized by tuning the hyperparameters.
We report the accuracy of the surrogate labels obtained for the best set of hyperparameters that minimizes this objective.

%% file: 5.experiments.tex
\begin{table*}[tb]
\centering
\caption{Datasets and the corresponding split-models used in our experiments.}
\scalebox{0.85}{
\begin{tabular}{|c|c|c|c| c|c|}
\hline
\multirow{2}{*}{Dataset}&\multicolumn{2}{c|}{Config-1}&\multicolumn{2}{c|}{Config-2}&\multirow{2}{*}{$g'$}\\
\cline{2-5}
& \textbf{$f$} & \textbf{$g$} & \textbf{$f$} & \textbf{$g$}&\\
\hline
MNIST            & $Conv\times4$ & $FC\times2$ &  $Conv\times3$ & $Conv-FC\times2$&$FC\times3$         \\
FashionMNIST     & $Conv\times4$ & $FC\times2$ & $Conv\times3$ & $Conv-FC\times2$&$FC\times3$\\
CIFAR-10        & $Conv-Res\times3$ & $FC\times2$ & $Conv-Res\times2$ & $Res-FC\times2$&$FC\times3$\\
CIFAR-100        & $Conv-Res\times3$ & $FC\times2$ & $Conv-Res\times2$ & $Res-FC\times2$&$FC\times3$\\
Criteo           & $Emb-FC\times2$ & $FC\times2$ &$Emb-FC$ & $FC\times3$&$FC\times4$\\
\hline
\end{tabular}}
\label{table:datasets}
\end{table*}

\section{Experiments}
We evaluate ExPLoit with multiple datasets and model architectures to show that it can leak private labels with high accuracy, across different settings. We describe our experimental setup followed by the results in this section.

\subsection{Experimental Setup}
The datasets and the corresponding split-models ($f \circ g$) used in our evaluations are shown in Table~\ref{table:datasets}.

\textbf{Datasets:}  MNIST, FashionMNIST, CIFAR-10, and CIFAR-100 are computer vision datasets used to perform multi-class image classification. The Criteo dataset consists of conversion logs for online ad-clicks, with each entry consisting of 3 continuous, and 17 categorical features, along with a binary label indicating if the ad-click resulted in a purchase (conversion). 
Note that the Criteo dataset has a large class imbalance (~$90\%$ of the labels are 0's, and the rest are 1's).  

\textbf{Models:} We use a 4-layer convolutional neural network for MNIST and FashionMNIST and a 21-layer ResNet model for CIFAR-10 and CIFAR-100. The model for the conversion prediction task (Criteo) consists of a learnable embedding layer (to handle categorical features) followed by four fully connected (FC) layers. All the models are split into two sub-models $f$ (input-owner's model) and $g$ (label-owner's model), which are jointly trained using split learning. The layer at which the model is split is referred to as the \emph{cut-layer}. To test the sensitivity of our attack to the cut-layer, we perform experiments with two different configurations: Config-1 and Config-2, which splits the model at different points as shown in Table~\ref{table:datasets}. The label-owner's model $g$ only consists of FC layers in Config-1. In contrast, the $g$ model in Config-2 is larger and consists of both convolutional (Conv) and FC layers. The vision models are trained for 10 epochs and the CVR model for 5 epochs using the Adam optimizer with a learning rate of 0.001. We evaluate the label leakage attacks after each training epoch.

\textbf{Attack parameters:} We assume that the architecture of the label owner's model $g$ is not known to the input owner. Thus, we use a 3-layer fully connected DNN (FC[128-64-10] for image classification and FC[32-32-10] for Criteo) as the surrogate model $g'$. We set $N_{iter}=500$, and the learning rate range to $[10^{-5}, 10^{-4}]$ for $\eta_{g'}$ and $[10^{-2}, 10^{-1}]$ for $\eta_{\hat{y}}$. The range for $\lambda_{ce}$ and $\lambda_p$ is set to $[0.1,3]$.  We carry out the ExPLoit attack for each training epoch of split learning. 

\textbf{Evaluation Metric:} ExPLoit groups inputs that belong to the same class. However, the true class label corresponding to each group remains unknown.  We report the clustering accuracy~\cite{clustering_accuracy} obtained with the best hyperparameters (corresponding to the lowest gradient loss) in our results.

\subsection{Results}\label{sec:results}

\begin{figure*}[tb]
	\centering
    \centerline{\epsfig{file=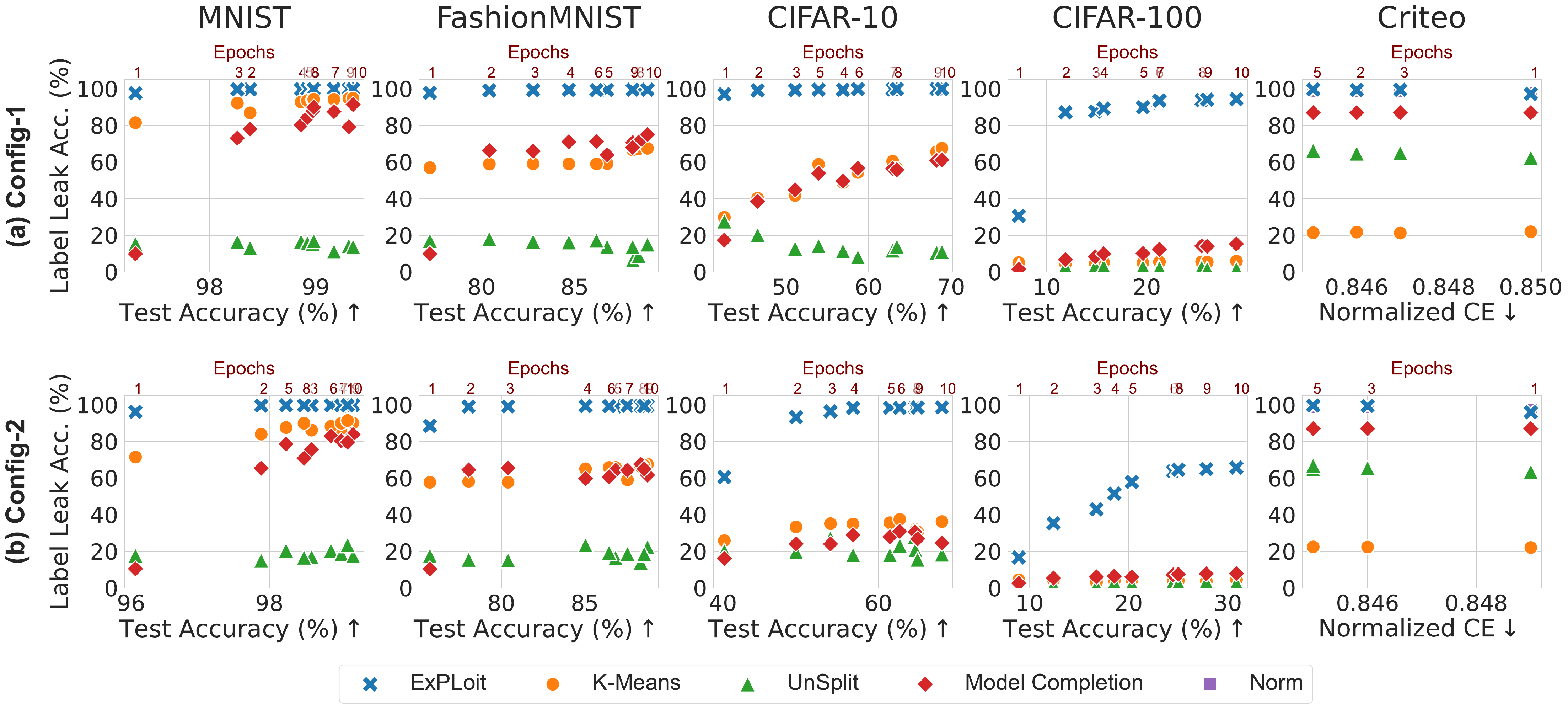, width=\textwidth}}
	\caption{Results comparing \emph{ExPLoit} with the K-means baseline and prior works (Unsplit, Model Completion and Norm-based attacks) for two configurations of the split model: (a) Config-1 and (b) Config-2 (see Table~\ref{table:datasets}). ExPLoit significantly outperforms prior works and can leak private labels with a near-perfect accuracy of up to $\uptox\%$.} 
    \label{fig:attack_results}
    \vspace{-0.2in}
\end{figure*}

We plot label leakage accuracy for ExPLoit and other label leakage attacks and compare it with the test accuracy/normalized cross entropy\footnote{We use normalized cross-entropy (NCE) instead of test accuracy to measure the model performance for the Criteo dataset as it has a high class imbalance. A lower value of NCE indicates better performance.} of the split-model, across various datasets and model configurations in Fig.~\ref{fig:attack_results}. ExPLoit achieves near-perfect label leakage accuracy for most datasets ($\uptox\%$ for CIFAR-10) and significantly outperforms all prior works. We provide explanations for the attack sensitivity to various parameters like dataset, split-model training epoch, cut layer, and a detailed comparison with recent prior works below. We use the accuracy numbers from Config-1 (Fig.~\ref{fig:attack_results}a) , Epoch-10 to discuss the results, unless specified otherwise.

\textbf{Sensitivity to Split-model Training Epoch:} Our results show that the efficacy of ExPLoit improves when attacking the later epochs of the split-model training. For instance, the label leakage accuracy of ExPLoit for CIFAR-100 is just $30.65\%$ for Epoch-1 and improves to $94.4\%$ for Epoch-10. The reason for this trend is because our attack approximates the label owner's model $g$ using a fixed surrogate model $g'$. However, in reality, $g$ is not fixed, and changes as training progresses. The rate of this change is smaller for the latter epochs. Consequently, our ability to approximate $g$ with a fixed surrogate model improves for the later epochs, which improves our attack's efficacy. While ExPLoit already achieves near perfect accuracy for most datasets with just 10 epochs of split-model training, we expect this accuracy to improve further as the split-model is trained for a greater number of epochs.

\textbf{Sensitivity to Datasets:} ExPLoit is more effective for datasets with lower input dimensionality and fewer classes. For instance, ExPLoit has a label leakage accuracy of $99.72\%$ for MNIST, whereas the attack accuracy drops to $94.38\%$ for CIFAR-100. This is because, for CIFAR-100 our attack has a higher number of  number of learnable surrogate parameters compared to MNIST (due to increase in input size and number of output classes), which makes learning them harder.  

\textbf{Sensitivity to Cut layer and Model Architecture:} We evaluate our attack on two split network configurations: Config-1 and Config-2, which represent two different choices of the cut layer. ExPLoit achieves a higher label-leakage accuracy for Config-1 compared to Config-2. This difference in performance is most pronounced in case of CIFAR-100, where our attack produces a label leakage accuracy of $94.38\%$ for Config-1 and $65.74\%$ for Config-2. The reason for  this discrepancy is two-fold. First, $g$ is larger for Config-2, compared to Config-1, which makes it harder to approximate with a surrogate model $g'$ for Config-2. Second, our $g'$ model is architecturally similar to the $g$ model in Config-1 as both these models use FC layers, while the $g$ models in Conv-2 uses FC and Conv layers. Thus, the efficacy of our attack reduces when $g$ is larger and has a dissimilar model architecture compared to $g'$.

\textbf{Comparisons with Baseline and Prior Work:} We compare the performance of ExPLoit against an unsupervised learning baseline (K-Means Clustering) and three recent attacks: Unsplit, Model Completion and Norm based attack. The experimental setup for these prior works is described in Appendix~\ref{app:evaluation_prior_work}.

\emph{K-Means Attack:} Through the split-learning process, the label-owner is able to generate embeddings $z_i=f(x_i)$, for each input $x_i$. One way to estimate the private labels is by using unsupervised learning to group these embeddings. We perform K-Means clustering using the embeddings $\{z_i\}$ and report the resulting label leakage accuracy in Fig.~\ref{fig:attack_results}. The efficacy of the attack improves with the quality of embeddings. Consequently, the attack performs well for simpler datasets like MNIST and FashionMNIST, which learn good embeddings with relatively few training epochs. The attack accuracy accuracy also improves for later epochs as the model $f$ learns better embeddings as training progresses. 

\emph{Unsplit Attack~\cite{unsplit}:} The UnSplit attack uses the gradient matching loss to learn the private labels. The authors of~\cite{unsplit} showed that this technique is effective only when $g$ is a single layer network and does not work for multi-layer networks. Consistent with their results, our experiments with the UnSplit attack also showed very low efficacy ($10.99\%$ attack accuracy for CIFAR-10, which is comparable to a random guess). 

\emph{Model Completion Attack\cite{lia_vfl}:} This attack proposes to train a surrogate model $g'$ using semi-supervised learning to predict the private labels corresponding to the inputs. Similar to the K-Means attack, the efficacy of this attack depends on the quality of emebddings produced by $f$. Consequently, this attack works well for simpler datasets, providing $91.46\%$ accuracy for MNIST, the accuracy is much lower at $61.3\%$ for CIFAR-10.

\emph{Norm-based attack\cite{label_leakage} :} The norm-based attack uses gradient norm to predict the labels in imbalanced binary classification problems. Evaluations on the Criteo dataset shows that this attack can achieve high accuracy (comparable with ExPLoit). The leakage accuracy also improves for the later epochs of the split-model training as the difference in gradient norms becomes more pronounced.

ExPLoit significantly outperforms all prior works, providing a near-perfect label leakage accuracy for most datasets. E.g. ExPLoit achieves a label-leakage accuracy of $\uptox\%$ for CIFAR-10, which is $32.38\%$ higher than the next best attack. This difference is even more pronounced when the attack is carried out at an earlier training epoch ($67.2\%$ higher accuracy compared to next best attack at Epoch-1 for CIFAR-10). This is because, unlike prior works, the efficacy of ExPLoit does not depend on the quality of embeddings produced by $f$. The efficacy of our attack demonstrates that split learning offers negligible privacy benefits for the label owner.

%% file: 6.gradient_noise.tex
\begin{figure*}[t]
	\centering
    \centerline{\epsfig{file=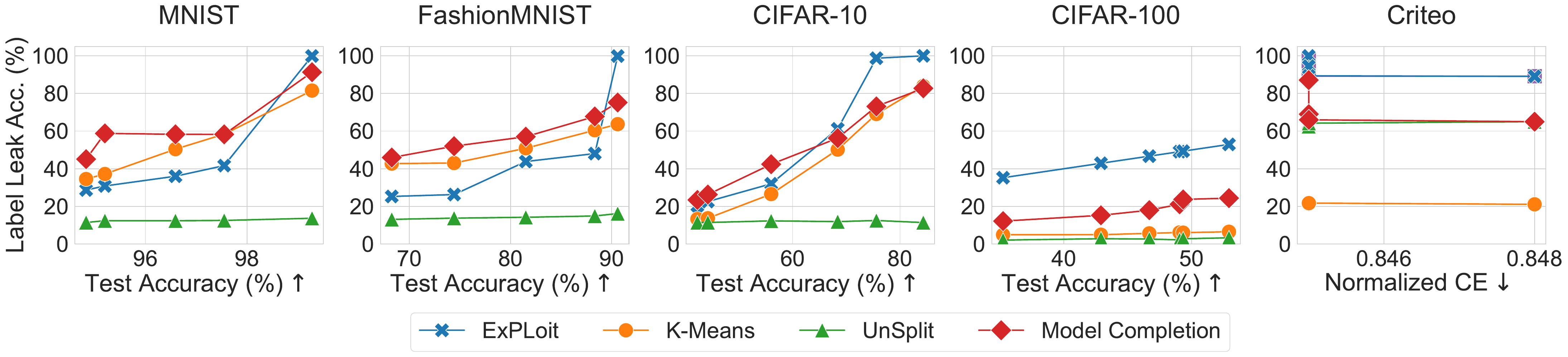, width=\textwidth}}
    \vspace{-0.1in}
	\caption{Results showing the utility-privacy trade-off for the gradient noise defense.} 
    \label{fig:defense_results}
    \vspace{-0.1in}
\end{figure*}

\section{Gradient Noise Defense}
ExPLoit uses the gradient information obtained from the label owner during split learning to leak the private labels. One way to defend against ExPLoit is by perturbing the loss gradients $\nabla_zL_i$ with noise as shown in Eqn.~\ref{eq:grad_noise}. 
\begin{align} \label{eq:grad_noise}
\nabla_z\hat{L}_i = \nabla_zL_i + \eta,\ \ where\ \eta \sim \mathcal{N}(0,\sigma)
\end{align}
The label owner can transmit these noisy gradients $\nabla_z\hat{L}_i$ to the input owner to perform split learning. Adding noise prevents the input owner from having reliable access to the true gradients. This reduces the efficacy of ExPLoit, providing better privacy to the label owner. On the other hand, noisy gradients are detrimental to training the split model and results in lower accuracy, thus impacting utility. To evaluate the utility-privacy trade-off offered by this defense, we perform split learning with different amounts of gradient noise by sweeping $\sigma$ in Eqn.~\ref{eq:grad_noise}. For each value of $\sigma$, we train the split-model till convergence and report the test and label leakage accuracy with ExPLoit.
Since the gradients obtained during split learning are noisy, optimizing the hyperparameters of our attack using the gradient loss objective is not optimal. Instead we tune the hyperparameters using $L_{ExP}(\lambda_{ce}=1, \lambda_p=1)$ as the optimization objective.

The utility-privacy trade-off with gradient noise defense for ExPLoit is shown in Fig.~\ref{fig:defense_results}. 
Against the ExPLoit attack, this defense provides a better utility-privacy trade-off for lower dimensional datasets like Criteo, MNIST and FashionMNIST. For instance, gradient noise degrades the label leakage accuracy for MNIST by $77\%$ with only a $4\%$ reduction of test accuracy. In contrast, for CIFAR-10, a $79\%$ reduction in label leakage accuracy incurs a $38\%$ reduction in test accuracy. This discrepancy is because adding gradient noise hampers the quality of the input owner's model $f$. Simpler datasets are more resilient to this degradation, whereas more complex datasets like CIFAR-10 and CIFAR-100 are impacted more if $f$ is not trained properly. Thus, while gradient noise might be suitable for simpler datasets, it may not be practical for more complex datasets. 

Additionally, we evaluate the gradient noise defense against other attacks. While the gradients are not directly used by the K-Means and model completion attacks, gradient noise reduces the quality of embeddings learnt by $f$, which degrades the efficacy of these attacks. We find that these attacks are more resilient to gradient noise for simpler datasets (MNIST and FashionMNIST) compared to ExPLoit as the quality of the embeddings does not degrade significantly. However, ExpLoit performs better than these two attacks for more complext datasets like CIFAR-100. The UnSplit attack continues to provide no benefit in the presence of the gradient noise defense.

%% file: 7.conclusion.tex
\section{Conclusion}
Split learning has been proposed as a method to train a model on vertically split data while keeping the data private. We investigate the privacy properties of two-party split learning by proposing \emph{ExPLoit} -- a label-leakage attack that allows an adversarial input owner to learn the label owner's private labels during split learning. Our key insight is that the attack can be framed as a learning problem by substituting the unknown parameters of the label owner with learnable surrogate parameters. We use the gradient data collected during split learning and a novel loss function to train these surrogate parameters. Our evaluations on several image-classification tasks and a converstion prediction task show that ExPLoit can leak private labels with near-perfect accuracy of up to $\uptox \%$, proving that split learning provides a negligible amount of label privacy. ExPLoit also outperforms recent prior works, offering up to $67.2 \%$ improvement in label leakage accuracy. We also evaluate gradient noise as a defense to improve label privacy. While this provides a reasonable defense for simpler datasets, we find that the utility-privacy tradeoff of this technique is unfavorable for more complex datasets. Our findings in this work underscore the need for better techniques to perform vertical federated learning.


%% file: acknowledgements.tex
\section{Acknowledgements}
We thank Anish Saxena, Neal Mangaokar, Ousmane Dia and our colleagues from the Memory Systems Lab for their feedback. This work was partially supported by a gift from Facebook. We also gratefully acknowledge the support of NVIDIA Corporation with the donation of the Titan V GPU used for this research.

%% file: 8.appendix.tex
\newpage

\appendix

\section{Other Privacy-Preserving Training Techniques for Vertical Federated Learning}\label{app:other_related}

In addition to split learning, several methods have been proposed to train a model on vertically partitioned private data. These methods can broadly be classified into three categories: 1. Differential Privacy (DP) 2. Multi-Party Compute (MPC) and 3. Trusted Execution Environment (TEE). We discuss solutions in each category and describe their limitations.

\textbf{Label Differential Privacy:} Differential privacy~\cite{dp} is a principled system for training on a private database that restricts the influence of any single entry of the database on the outcome by adding noise to a query's response. A recent work~\cite{labeldp} proposed \emph{Label Differential Privacy (LDP)} to train a model on vertically partitioned data with sensitive labels. LDP relies on a randomized response algorithm to provide a noisy version of the labels to the input owner by defining a probability distribution over the class labels as follows:   
\begin{align} \label{eq:ldp}
    \operatorname{Pr}[\tilde{y}=\hat{y}]= \begin{cases}\frac{e^{\varepsilon}}{e^{\varepsilon}+K-1} & \text { for } \hat{y}=y \\ \frac{1}{e^{\varepsilon}+K-1} & \text { otherwise }\end{cases}
\end{align}

The label owner uses the noisy labels sampled from this distribution and the input data to train a model. To prevent the model from overfitting on the incorrect labels, the authors propose using the Mixup technique~\cite{mixup} to train the model, which provides resilience to label noise.
One drawback of this technique is that it allows the input owner to have complete ownership of the model. In contrast, split learning enables the input and label owners to jointly own the model, which might be desirable if the label owner wants to exercise control over the usage of the model.

\textbf{Multi-Party Compute (MPC):} Several works\cite{mpc1,mpc2,secureml} have proposed using cryptographic techniques to enable private computations over distributed data held by multiple parties. These works use a combination of cryptographic primitives such as oblivious transfer~\cite{ot1,ot2,ot3}, garbled circuits~\cite{gc}, secret sharing~\cite{secret} and homomorphic encryption~\cite{paillier1999public} to train the model. Unfortunately, these methods have significant computational overheads and require multiple rounds of communication between the parties involved. Consequently, even training a simple 2-layer network incurs a $30\times$ overhead~\cite{secureml} compared to training without privacy, making it impractical for training larger networks. 

\textbf{Trusted Execution Environment (TEE):} Trusted Execution Environments use hardware enclaves to enable remote computations with confidentiality and integrity. A centrally hosted TEE can be used to train a model on distributed data. The data owners can communicate data securely over an encrypted channel to the trusted enclave. Training is performed while ensuring data confidentiality, and the resulting model is transmitted securely to the data owners. Unfortunately, TEEs have slow memory due to the overheads associated with encryption and integrity checks~\cite{sgx_mem1,sgx_mem2}. Moreover, commercially available TEEs such as Intel SGX~\cite{sgx} and Arm Trustzone~\cite{trustzone} are CPU-based and offer less parallelism compared to GPUs. The combination of these two factors results in orders of magnitude~\cite{sgx_training} increase in training times of models. Additionally, this solution requires specialized hardware, which adds to the cost of implementation.

\begin{table*}[htb]
\centering
\caption{Ablation Study showing the label leakage accuracy of ExPLoit when the two regularization terms: Label Prior Regularization (LPR) and Cross Entropy Regularization (CER), are not used.}
\begin{tabular}{|c|c|c|c|c|}
\hline
Dataset      & Original (\%) & No LPR (\%)& No CER (\%)& No LPR, CER (\%)\\
\hline
MNIST        & 99.82    & 68.29 & 31.43 & 17.95       \\
FMNIST & 99.84    & 69.78 & 59.31 & 27.83       \\
CIFAR-10     & 99.96   & 99.28  & 99.35 & 99.84       \\
CIFAR-100    & 94.38    & 60.78 & 17.04 & 20.38      \\
Criteo       & 99.68   & 99.65 & 99.87 & 97.75     \\
\hline
\end{tabular}
\label{table:ablation}
\end{table*}

\section{Limitations}\label{app:limitations}
Our proposed ExPLoit Attack uses the gradient information obtained during split learning to leak the private labels. We assume that the input owner has knowledge of the number of classes and the distribution of the labels over these classes (prior information) to develop our loss function (Eqn.~\ref{eq:loss}). If the attacker is completely unaware of the downstream classification task, this information could be hard to estimate, making our attack less effective (see Appendix~\ref{app:ablation}). However, we argue that it is rare for the input owner (attacker) to be completely unaware of the downstream classification task. Even if the label prior is unknown, knowledge of the task in itself might be sufficient to make an educated guess about the label prior. For instance, in the case of conversion prediction, the average conversion rate for online advertising is publicly available~\cite{cvr}. In the case of disease prediction, the prevalence rate of a disease is often known and can be used as the label prior. Developing attacks that do not require label prior information is an interesting avenue of exploration for future work.

\section{Ablation Study}\label{app:ablation}
Our attack replaces the unknown labels and model of the label owner with surrogate labels and uses the gradient information obtained during split learning to train these parameters.
In addition to matching the surrogate gradients obtained during ``replay" split learning with the original gradients, our loss function consists of two regularization terms as shown in Eqn.~\ref{eq:loss2}. 
\begin{equation}\label{eq:loss2}
\begin{split}
L_{ExP} =& \mathbb{E}\Big[ \left\|\nabla_{z}L_i - \nabla_{z}L'_i\right\|_2\Big] +\\ 
&\lambda_{ce}\cdot\mathbb{E}\Big[H(y'_i, p'_i)/H(P_y)\Big] + \lambda_p \cdot \mathcal{D}_{KL}(P_{y}\| P_{y'})
\end{split}
\end{equation}

The cross-entropy regularization (CER) term $\mathbb{E}\Big[H(y'_i, p'_i)/H(P_y)\Big]$ achieves the dual objective of minimizing the entropy of the individual surrogate labels and improving the accuracy of the surrogate model ($g'$). The label prior regularization (LPR) term $\mathcal{D}_{KL}(P_{y}\| P_{y'}$) tries to match the distribution of the surrogate labels with the label prior. We conduct an ablation study to understand the importance of the two regularization terms by carrying out ExPLoit without using LPR, without using CER, and without using both LPR and CER. The resuls of this study are shown in Table~\ref{table:ablation}. As expected, we find that there is a degradation in accuracy when regularization terms are not used. CER seems to be more important compared to LPR as the degradation is higher when CER is not used. For CIFAR-10 and Criteo, the regularization terms seem to matter less as the accuracy is high even when we disable both regularization terms.

\section{Experimental Setup for Prior Works} \label{app:evaluation_prior_work}
We describe the experimental setup and evaluation methodology for the prior works used in our experiments. The Unsplit and Model Completion attacks both require a surrogate model. To have a fair comparison, we use the same surrogate model as ExPLoit (see Table~\ref{table:datasets}) for both of these attacks.

\textbf{UnSplit Attack:} The UnSplit attack~\cite{unsplit} aims to learn the surrogate labels $\{y'\}$ and model parameters $\theta_{g'}$ by minimizing the mean square error loss between the original and the surrogate gradients using the following objective: $\min_{\theta_g', \{y'_i\}} MSE(\nabla_{z}L'_i, \nabla_{z}L_i)$. We use an Adam optimizer with a learning rate of 0.001 and train for 50 epochs.

\textbf{Model Completion Attack:} Model completion attack~\cite{lia_vfl} trains the surrogate model using semi-supervised learning by using a small number of labeled embeddings $\mathcal{D}^l = \{z^l_i, y^l_i\}$ and  unlabeled embeddings  $\mathcal{D} = \{z_i\}$. We assume that the attacker has 4 labeled examples per class for the vision datasets and 50 examples per class for the Critio dataset. We use the parameters from the original paper ~\cite{lia_vfl} to perform semi-supervised learning. We set temperature T=0.8 for sharpening the predictions, $\lambda_u=50$ as the weight for the loss on the unlabeled dataset and $\alpha=0.5$ for MixUp. We train all the models for 100 epochs and report the accuracy by using the predictions of this trained model.

\begin{figure*}[htb]
	\centering
    \centerline{\epsfig{file=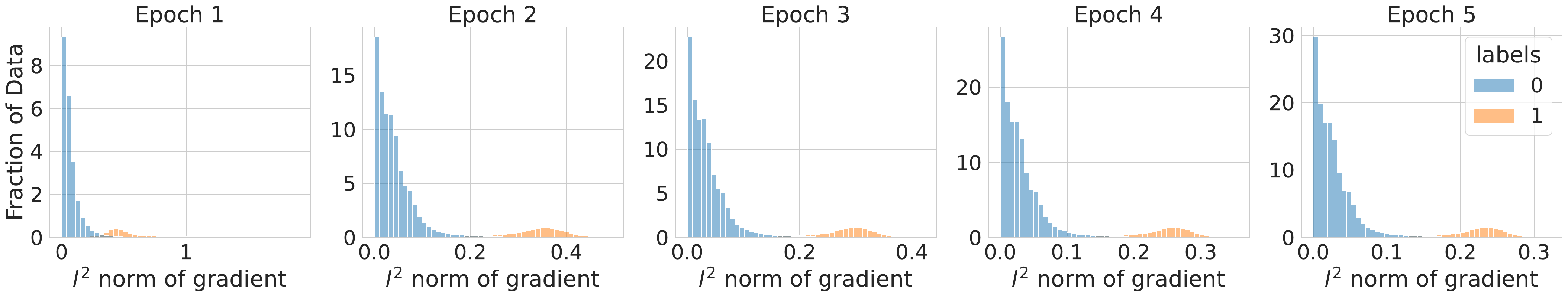, width=\textwidth}}
	\caption{Distribution of gradient norms $\|\nabla_z L\|_2$ for the conversion prediction task with Criteo dataset. Positive classes are infrequent and typically produce higher gradient norms.}
	\label{fig:norm_attackk}
\end{figure*}

\textbf{Norm-Based Attack:} The norm-based attack uses the difference in the magnitude of gradient norms in imbalanced binary datasets to predict the private labels. This difference can be seen clearly from Fig.~\ref{fig:norm_attackk}, which shows the distribution of gradient norms $\|\nabla_z L\|_2$ obtained during split learning for different epochs of the conversion prediction model trained on the Criteo dataset. The norm-based attack exploits this difference in the gradient norms and uses it to infer the private labels in split learning. This attack uses a threshold T to classify the examples into positive and negative classes as follows:
\begin{align} \label{eq:norm_attack}
    y'= \begin{cases}1 & \|\nabla L_z\|_2 > T  \\ 0 & \text { otherwise }\end{cases}
\end{align}
We sweep the value of $T$ and pick a threshold that gives the best accuracy value to report the Norm-based attack results in Section~\ref{sec:results}. Note that in a real attack setting, the adversary does not have the ability to check the accuracy for different values of $T$. Our goal in doing so is to understand the best possible accuracy that can be obtained with the Norm-based attack.

\section{Input Privacy Attacks and Defenses in Split Learning}\label{app:input_privacy_split_learning}
Recent works have proposed attacks to break input privacy in split learning. The goal of these attacks (a.k.a model inversion attack) is for an adversarial label owner to recover the private inputs $\{x_i\}$ of the input owner using the embedding information $\{z_i\}$ obtained during split learning. A recent work~\cite{input_leakage_1d} showed that, for simple 1-d time-series signals, the embedding data obtained in split learning might not preserve privacy as it has a high distance correlation with the original input data. Model inversion attacks have also been demonstrated on split learning with more complex datasets in the image domain~\cite{titcombe2021practical}. To carry out the attack, the adversary uses examples from the input data distribution $\{x'_i\}$ to query the input owner's model and generate embeddings $z'_i = f(x'_i)$. The input and embedding data can be used to train an inversion model $f_{inv}$ that maps the embedding to the input: $\mathcal{Z} \rightarrow \mathcal{X}$. This inversion model can be used to reconstruct the input data using the embeddings during the attack. Note that such attacks require access to the examples from the input data distribution and black-box query access to the input owner's model. In contrast, our label leakage attack does not require black-box access to the label owner's model or access to the ground truth label data. ~\cite{input_leakage_1d} also proposes using additive noise to perturb the embedding to defend against such attacks. This is similar in spirit to our work, which uses gradient noise to deter label leakage attacks.